\documentclass[preprint,12pt]{elsarticle}

\usepackage{amssymb}
\usepackage{xcolor}
\usepackage[T1]{fontenc} 
\usepackage[utf8]{inputenc}
\usepackage{hyperref}
\usepackage{graphicx}
\usepackage{bm}
\usepackage{amsmath,amssymb}

\usepackage{ragged2e}
\usepackage{fancyhdr}
\usepackage{xspace}

\journal{Physics of the Dark Universe}

\begin{document}

\begin{frontmatter}



\title{Beyond $\Lambda$CDM with $f(z)$CDM: criticalities and solutions of Pad\'e Cosmography}


\author[1]{A. Turmina Petreca}
\affiliation[1]{organization={Instituto de Física, Universidade de São Paulo},
            addressline={ Rua do Matão 1371}, 
            city={São Paulo},
            postcode={CEP 05508-090}, 
            state={São Paulo},
            country={Brazil}}

\author[2,3]{M. Benetti}
\author[2,3,4]{S. Capozziello}

\affiliation[2]{organization={Scuola Superiore Meridionale},
            addressline={, Largo S. Marcellino 10}, 
            city={Napoli},
            postcode={I-80138}, 
            country={Italy}}
            
\affiliation[3]{organization={Istituto Nazionale di Fisica Nucleare (INFN), sez. di Napoli},
            addressline={Via Cinthia 9}, 
            city={Napoli},
            postcode={I-80126}, 
            country={Italy}}
            
\affiliation[4]{organization={Dipartimento di Fisica  ``E. Pancini", Università di Napoli  ``Federico II"},
            addressline={Via Cinthia 9}, 
            city={Napoli},
            postcode={I-80126}, 
            country={Italy}}

\begin{abstract}
Recently, cosmography  emerged as a valuable tool to effectively describe the vast amount of astrophysical observations without relying on a specific  cosmological model. Its model-independent nature ensures a faithful representation of  data, free from theoretical biases. Indeed, the commonly assumed fiducial model, the $\Lambda$CDM, shows some shortcomings and tensions between data at late and early times that need to be further investigated. 
In this paper, we explore an extension of the standard cosmological model by adopting the $f(z)$CDM approach, where $f(z)$ represents the cosmographic series  characterizing the evolution of  recent universe driven by dark energy. 
To construct $f(z)$, we take into account the Pad\'e series, since this rational polynomial approximation offers a better convergence at high redshifts than the standard Taylor series expansion. Several orders of such an approximant have been proposed in previous works, here we want to answer the questions: What is the impact of  the cosmographic series choice on the parameter constraints? Which series is the best for the analysis? So, we analyse the most promising ones by identifying which order is preferred in terms of stability and goodness of fit. Theoretical predictions of the $f(z)$CDM model are obtained by the Boltzmann solver code and the posterior distributions of the cosmological and cosmographic parameters are constrained by a Monte Carlo Markov Chains analysis. We consider a joint data set of cosmic microwave background temperature measurements from the Planck collaboration, type Ia supernovae data from the latest Pantheon+ sample, baryonic acoustic oscillations and cosmic chronometers data. 
In  conclusions, we state which series can be used when only late time data are used, while which orders has to be considered in order to achieve the necessary stability when large redshifts are considered.
\end{abstract}


\end{frontmatter}


\section{Introduction}
\label{sec:intro}

The current standard cosmological model has proven to be able to overcome numerous observational challenges, providing a remarkably successful framework to explain both the primordial \cite{WMAP:2012nax,Planck:2018vyg} and large-scale structure  evolution of the universe \cite{SupernovaSearchTeam:1998fmf, Riess:2019cxk, Riess:2021jrx, Scolnic:2021amr, Brout:2022vxf}.  This widely accepted model is not without  shortcomings \cite{DiValentino:2020zio, Abdalla:2022yfr, DiValentino:2022fjm, Lopez-Corredoira:2017rqn}, and  most of its intriguing aspects lie on the nature of its dominant constituents: dark matter \cite{Ostriker:1973uit, Arbey:2021gdg} and dark energy \cite{Peebles:2002gy, Copeland:2006wr}. Both these exotic components have been postulated to explain unresolved observational issues, such as observed gravitational effects at galactic and extra-galactic scales as well as the current accelerated expansion of the universe. 
However,  there is  no final answer on the fundamental nature of dark matter despite of the fact that such an ingredient manifests itself  for the clustering of structures \cite{Capozziello:2011xp}.
Furthermore, dark energy constitutes about $70\%$ of the today energy density of the universe \cite{Planck:2018vyg}, and its influence  has been deduced from astrophysical observations, i.e. measurements of distant supernovae which  revealed an accelerated expansion of the Hubble flow \cite{SupernovaSearchTeam:1998fmf, Riess:2019cxk, Riess:2021jrx}. 
Despite its crucial role, also the fundamental nature of dark energy is still unknown, leaving various theories to emerge about it.  For example  scalar fields  \cite{Carroll:2000fy,Sahni:1999gb} or  modified gravity \cite{Capozziello:2011et,Capozziello:2007ec,Bamba:2012cp,Nojiri:2017ncd} can be invoked as mechanisms to source the accelerated expansion. Furthermore,  the puzzle  of   dark energy evolution, as a constant or a scale-dependent ingredient, is still matter of debate \cite{Carroll:2003st,Bargiacchi:2021hdp,Gonzalez:2021ojp}. The answers to these questions have far-reaching implications for our understanding of the universe and its ultimate destiny. 

Recently, a further significant challenge for the cosmological standard model stems out from  tensions that  emerged between the $\Lambda$CDM predictions and the observations of  current universe \cite{Schoneberg:2021qvd,DiValentino:2020zio,DiValentino:2021izs}. On the one hand, measurements of the cosmic microwave background radiation \cite{Planck:2018vyg}, primordial nucleosynthesis \cite{Cyburt:2004yc}, and the large-scale distribution of matter \cite{Anderson:2012sa} strongly support the $\Lambda$CDM model framework. On the other hand, observations of the current cosmic expansion rate, inferred from measurements of the Hubble constant using SNe \cite{Riess:2021jrx} show a discrepancy at more than 4-$\sigma$ when compared to the value predicted by the standard cosmological model \cite{SupernovaSearchTeam:1998fmf, Riess:2019cxk}. This tension, known as the $H_0$ tension, has emerged in last decade and it is a hot topic that deserves both a thorough investigation of possible systematics in the data \cite{HST:2000azd,Riess:2023bfx}, and the analysis of extensions or modifications of the $\Lambda$CDM model \cite{Schoneberg:2021qvd,DiValentino:2020zio,DiValentino:2021izs,Benetti:2019lxu,  Borges:2023xwx,Benetti:2021div,Salzano:2021zxk,Benetti:2017gvm} and the pillars on which it is based, such as the homogeneity at large scale \cite{Mustapha:1997fjz, Goncalves:2017dzs, Goncalves:2018sxa, Gonzalez:2021ojp, Andrade:2022imy} and general relativity as the basic theory of the model \cite{Ishak:2005zs,Harko:2020ibn, Capozziello:2012eu, Bajardi:2022ocw, Abdalla:2022yfr, Benetti:2020hxp, Benetti:2018zhv}.

In addition, degeneracies among cosmological models are a further issue in finding a unique solution within the $\Lambda$CDM context. Indeed, different extensions of the standard cosmological model may produce similar observational results, thus leaving the challenge of figuring out what new physics or mechanism needs to be considered \cite{Efstathiou:1998xx,vonMarttens:2019ixw, Schoneberg:2021qvd,Abdalla:2022yfr}. Identifying degeneracies and distinguishing among the various scenarios are crucial steps in refining the model predictions and improving our understanding of the fundamental properties of the universe.

Some of the challenges of the current cosmological model can be approached by exploring new tools and techniques, e.g. exploring the effects of dark energy without explicitly assuming a specific cosmological framework \cite{Shafieloo:2005nd,Sahni:2008xx,Daly:2003iy,LHuillier:2017ani}. Cosmography, a well-known technique used in  data analysis, offers a promising application in this field \cite{Busti:2015xqa,Visser:2004bf,Dunsby:2015ers,Capozziello:2019cav,Yang:2019vgk, Aviles:2012ir, Aviles:2012ay, Aviles:2013nga, Aviles:2016wel}. By directly probing the large-scale structure of the universe through observational data, cosmography provides a model-independent way to map the effects of dark energy and to study its impact on cosmic expansion. The advantages of using cosmography are several, including the ability to circumvent parameter degeneracies in the cosmological models \cite{Capozziello:2019cav}. Characterizing the expansion of the universe directly from observational data can reduce the number of free parameters while improving data fitting and producing rigorous constraints on cosmological evolution.

However, cosmography is not without  issues. One of the main challenges is the convergence of the series expansion  \cite{Cattoen:2007sk,Capozziello:2020ctn, Gruber:2013wua, Capozziello:2019cav, Lobo:2020hcz}. In fact, several cosmographic methods are based on Taylor series expansions, which show convergence problems for $z>1$ \cite{Capozziello:2019cav, Lobo:2020hcz}. It is a significant limit to the accuracy and precision of cosmographic reconstructions, especially if higher redshifts have to be explored. 

To overcome the convergence limitations, an emerging alternative approach  is the use of Pad\'e rational approximant \cite{Pade, Aviles:2014rma, Mehrabi:2018oke, Rezaei:2017yyj, Wei:2013jya, Gruber:2013wua, Capozziello:2020ctn, Zhou:2016nik, Liu:2021oaj, Capozziello:2017ddd, Benetti:2019gmo, Dutta:2018vmq, Dutta:2019pio}. Unlike traditional Taylor expansions, Pad\'e allows for a better representation of functions with poles or singularities, making them better suited to handle the complexities often encountered in cosmological analyses. By using Pad\'e polynomial, the cosmography can be used at redshift higher than one, providing a more reliable framework for studying the evolution of the universe \cite{Benetti:2019gmo, Capozziello:2018jya,DAgostino:2023cgx}. 

Given the improvements in cosmographic techniques, new approaches  emerged in recent years.  Instead of considering the  universe expansion led by some scalar field or modified gravity with constant or scale-dependent equation of state, $w$, it is possible to parameterize, in the cosmological equations,  the $\Lambda$ constant/dark energy content  with a function $f(z)$, and either leave this function generic \cite{Dutta:2018vmq, Dutta:2019pio} or assign it some specific form \cite{Benetti:2019gmo}. Hence, the concept of  $f(z)$CDM model arises, i.e assuming a generic  universe expansion where the current epoch is  figured out  by cosmographic parameters constrained by observations. From the value of these parameters, it is possible then to extract  information about the behavior of the current expansion without assigning a priori a given cosmological model.

At this point, it is important to ask whether there is and, if so, what is the impact of the specific form of the $f(z)$ considered on the constrained values of the cosmographic parameters. Furthermore, given the variety of choices  within the same approximant, what is the best choice in terms of regularity of the function, and number of cosmographic parameters considered.  

In this paper, we aim to answer these questions by considering three sets of Pad\'e's series well studied in  literature. In addition,  we provide new constraints with the most up-to-date data on the $f(z)$CDM model, considering, in particular,  the stability of  $P_{21}$, $P_{22}$ and $P_{32}$ series, inferring the best choice for different cosmological issues.

The paper is organized as follows. We introduce cosmography and the improvement coming from Pad\'e's series in Sec. \ref{sec:theory-cosmography}. In particular,  the stability of the considered series is tested for various values of cosmographic parameters. We then describe the $f(z)$CDM  approach in Sec. \ref{sec:theoryfzCDM}. In Sec. \ref{sec:Method} we detail the method of analysis and the data used, as well as the results are described. Finally, in Sec. \ref{sec:Conslusions}, we discuss our results and draw our conclusions.

\section{The Pad\'e Cosmography}
\label{sec:theory-cosmography}

Let us discuss now  the cosmographic approach considering the rational Pad\'e  series \cite{Pade, Aviles:2014rma, Mehrabi:2018oke, Rezaei:2017yyj, Wei:2013jya, Gruber:2013wua, Capozziello:2020ctn, Zhou:2016nik, Liu:2021oaj, Capozziello:2017ddd, Benetti:2019gmo, Dutta:2018vmq, Dutta:2019pio}. In general, cosmography is built up starting from the Taylor expansion of the cosmic scale factor, $a(t)$, with $t$ being the cosmic time \cite{Capozziello:2019cav, Lobo:2020hcz}. The evolution of the universe is embodied in the scale factor because its dynamic is governed by the Friedman equations which give the expansion rate. Expanding it around the present epoch, $t_0$, we have:

\begin{equation}
a(t)=1+\sum_{k=1}^{\infty}\dfrac{1}{k!}\dfrac{d^k a}{dt^k}\bigg | _{t=t_0}(t-t_0)^k\,.
\label{eq:scale_factor}
\end{equation}
From this expression, it is possible to determine the Hubble parameter $H(t)$, and others cosmographic quantities, as deceleration $q(t)$, jerk $j(t)$, snap $s(t)$, lerk $l(t)$, pop $p(t)$, etc., up to the n\textit{-th} order. They are given by the following equations \cite{Capozziello:2019cav, Lobo:2020hcz}:
\begin{subequations}
\begin{equation}
H(t)\equiv \dfrac{1}{a}\dfrac{da}{dt} \ , \hspace{1cm} 
q(t)\equiv -\dfrac{1}{aH^2}\dfrac{d^2a}{dt^2}\ ,  \label{eq:hubble_decelerating} 
\end{equation}
\begin{equation}
j(t) \equiv \dfrac{1}{aH^3}\dfrac{d^3a}{dt^3}\ , \hspace{0.5cm}  
s(t)\equiv\dfrac{1}{aH^4}\dfrac{d^4a}{dt^4}\ , \label{eq:jerk_snap} 
\end{equation}
\begin{equation}
l(t) \equiv \dfrac{1}{aH^5}\dfrac{d^5a}{dt^5}\ , \hspace{0.5cm}  
p(t)\equiv\dfrac{1}{aH^6}\dfrac{d^6a}{dt^6}\ , \label{eq:lerk_pop} 
\end{equation}
\end{subequations}
These parameters take on physical significance as they are dependent on the derivatives of the Hubble parameter, and give us information about the universe evolution. In fact, from the sign of deceleration parameter, it is possible to infer if the universe is accelerating or decelerating, the sign of jerk indicates the time evolution of the universe acceleration, and the snap value gives information about the nature of  dark components (energy or matter) while higher order parameters can be included to refine the evolutionary behavior \cite{Lobo:2020hcz,Capozziello:2011tj}. 

In order to write the Hubble parameter in terms of the cosmographic ones and redshift, $z$, it is often considered a Taylor expansion as \cite{Capozziello:2011tj}:

\begin{equation}
    H(z) = H(0) + \frac{dH(z)}{dz} \bigg|_{z=0} z + \frac{d^2 H(z)}{dz^2} \bigg|_{z=0} z^2 + O(z^3)\,.
    \label{eq:taylor_H}
\end{equation}
 The derivatives of $H$(z) can be written with respect to the redshift. By replacing the derivatives of $H$(z) with respect to time and those of time in terms of redshift, we get \cite{Capozziello:2011tj}: 

\begin{equation}
\frac{dH}{dz}=H \frac{1+q}{1+z}
\ , \hspace{1cm} 
\frac{d^2 H}{dz^2} = H \frac{j-q^2}{(1+z)^2} \ , \hspace{1cm} ... \ 
\end{equation}

Eq.(\ref{eq:taylor_H}) can be rewritten in terms of cosmographic parameters, as \cite{Lobo:2020hcz, Capozziello:2020ctn}:

\begin{equation}
\begin{split}
     \frac{H(z)}{H_0} & = \{1 + (1+q_0) z + \frac{1}{2}(j_0-q_0^2) z^2-\frac{1}{6} [-3 q_0^2-3 q_0^3+j_0 (3+4q_0)+s_0]z^3\\
     &+\frac{1}{24}[-4 j_0^2+l_0-12 q_0^2-24 q_0^3 -15q_0^4+j_0(12+32 q_0 +25 q_0^2)+8 s_0 +7 q_0 s_0]z^4\\
     &+\frac{1}{120}(p_0+15 l_0+60 (s_0 + j_0 - j_0^2 + 4 j_0 q_0 + s_0 q_0^2 - q_0^2 - 3 q_0^3) - 15 s_0 j_0 + 11 l_0 q_0\\
     &+105 s_0 q_0 - 70 j_0^2 q_0 + 375 j_0 q_0^2 + 210 j_0 q_0^3 - 225 q_0^4 - 105 q_0^5) z^5 + O(z^6) \}\,.
\end{split}
\label{eq:Hz}
\end{equation}

Now, we have to consider the luminosity distance, $d_L$, which is  necessary  for calibrating cosmic distance scales, and then understanding the universe structure and evolution:

    \begin{equation}
        d_L = (1+z) \frac{c}{H_0} \int_{0}^{z} \frac{d z'}{H(z')}
        \label{eq:luminosity_distance}
    \end{equation}

  with $c$ being the speed of light. 
  Using the definitions given by  Eqs.(\ref{eq:hubble_decelerating})-(\ref{eq:lerk_pop}), knowing  $H(z)$ from Eq.(\ref{eq:Hz}),
  and the relation between redshift and scale factor, $z=\frac{a(t_0)}{a(t)} -1$, 
  it is possible to recast $d_L (z)$ in terms of cosmographic parameters \cite{Capozziello:2020ctn}:

\begin{multline}
    d_L(z) = \frac{cz}{H_0} \{1 + \frac{z}{2}(1-q_0) - \frac{z^2}{6}(1-q_0-3 q_0^2 + j_0) + \frac{z^3}{24}(2 - 2 q_0 - 15 q_0^2 - 15 q_0^3 + 5 j_0 + 10 q_0 j_0 + s_0)\\ 
    -z^4\left(\frac{1}{20}-\frac{9j_0}{40}+\frac{j_0^2}{12}-\frac{l_0}{120}+\frac{q_0}{20}-\frac{11q_0j_0}{12}+\frac{27q_0^2}{40}-\frac{7j_0q_0^2}{8}+\frac{11q_0^3}{8}+\frac{7q_0^4}{8}-\frac{11s_0}{120}-\frac{q_0s_0}{8} \right) \\
    + z^5 \left[-\frac{1}{20} - \frac{9 j_0}{40} + \frac{j_0^2}{12} - \frac{l_0}{120} + \frac{q_0}{20} - \frac{11 q_0 j_0}{12} + \frac{27 q_0^2}{40} - \frac{7 j_0 q_0^2}{8} + \frac{11 q_0^3}{8} + \frac{7 q_0^4}{8} - \frac{11 s_0}{120} - \frac{q_0 s_0}{8} \right] \\
    + O(z^6)\}
    \label{eq:taylor_dL}\,.
\end{multline}

Despite of the  wide use of this expression at low redshifts \cite{Capozziello:2020ctn}, the restricted convergence of the Taylor series makes this method poorly predictive for $z>1$ \cite{Lobo:2020hcz, Capozziello:2019cav}.  The problem can be partially alleviated by adopting the Pad\'e or Chebyshev rational polynomials \cite{Capozziello:2019cav,Gruber:2013wua, Capozziello:2017nbu} or logarithmic polynomial series \cite{Bargiacchi2021}. In this paper we consider Pad\'e approximant, defined as the ratio between two Taylor expansions of order $n$ and $m$ \cite{Capozziello:2019cav}
:
\begin{equation}
P_{nm}(z)=\dfrac{\displaystyle{\sum_{i=0}^{n}a_i z^i}}{1+\displaystyle{\sum_{j=1}^{m}b_j z^j}}\,.
\label{eq:Pnm_Pade}
\end{equation}
Such a rational approximation can alleviate  divergence problems of Taylor series for $z > 1$, since the denominator can stabilize the function, i.e., the series can be calibrated by choosing appropriate orders for a specific situation \cite{Capozziello:2019cav}. In general, we can note a worse stability in polynomials with the same order in numerator and denominator, and a better stability when the denominator order is lower than the numerator one \cite{Capozziello:2018jya}. Indeed, $P_{22}$ has been shown to have poor convergent behavior, compared to $P_{32}$ and  $P_{21}$ \cite{Capozziello:2020ctn}.  Therefore, in this paper, we  study these latter, but  consider also $P_{22}$ both to point out possibly other peculiarities and because it is widely used in literature.  We aim to better understand the application of  Pad\'e approximation in  cosmography  and compare   different orders of polynomials  to determine  goodness and criticality of the approach. The aim is understand how $\Lambda$CDM can be improved relaxing the strict requirement of $\Lambda$ constant according to the evolution of cosmic flow at various redshifts.
Specifically, we are going to consider the following Pad\'e series:

\begin{equation}
\label{eq:f_P21}
P_{21}(z) = \frac{P_{0}+P_{1}z+P_{2}z^{2}}{1+Q_{1}z}\,.
\end{equation}

\begin{equation}
\label{eq:f_P22}
P_{22}(z) = \frac{P_{0}+P_{1}z+P_{2}z^{2}}{1+Q_{1}z+Q_{2}z^{2}} \,.
\end{equation}

\begin{equation}
\label{eq:f_P32}
P_{32}(z) =\frac{P_{0}+P_{1}z+P_{2}z^{2}+P_{3}z^{3}}{1+Q_{1}z+Q_{2}z^{2}}.
\end{equation}
where the coefficients $P_0$, $P_1$, $P_2$, $P_3$, $Q_1$ and $Q_2$ can be determined by comparing these expressions (and their derivatives) with the Taylor expansion $f(z) = \sum_{i=0}^{\infty} c_i z^i$ (and its derivatives) calculated at $z=0$:
\begin{equation}
    P_{nm}(0) = f(0)
    \label{eq:0der}
\end{equation}
\begin{equation}
    P'_{nm}(0) = f'(0)
\end{equation}
\begin{equation*}
\vdots
\end{equation*}
\begin{equation}
    P_{nm}^{(n+m)}(0) = f^{(n+m)}(0)
    \label{eq:nmder}
\end{equation}
\begin{figure}[!]
    \centering
     \includegraphics[scale=0.4]{ 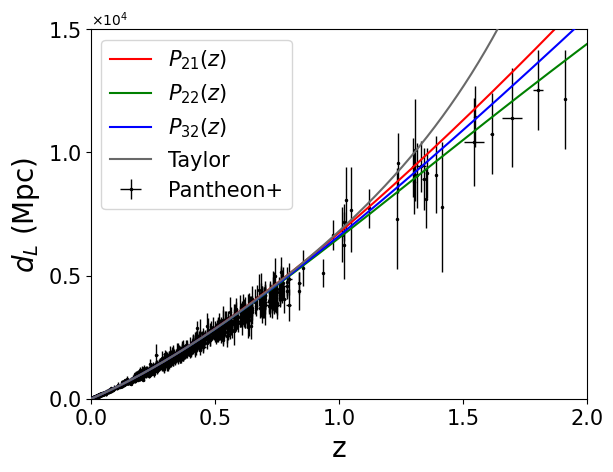}
    \includegraphics[scale=0.4]{ 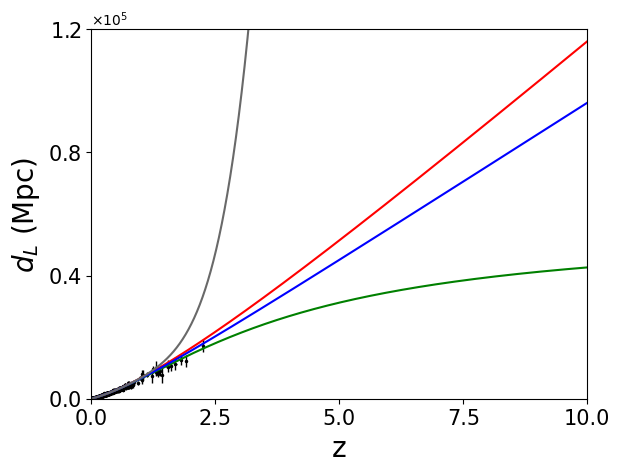}
    \includegraphics[scale=0.4]{ 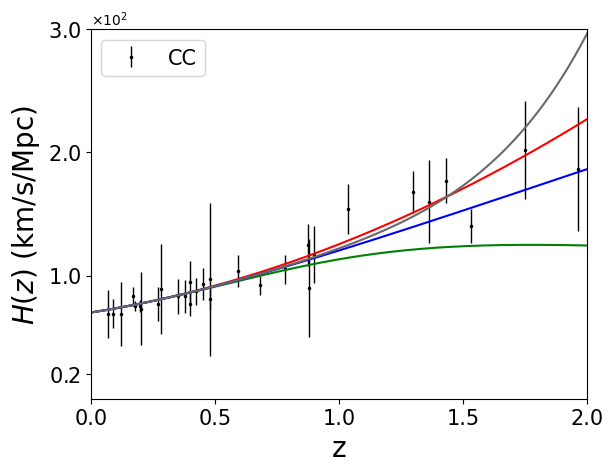}
    \includegraphics[scale=0.4]{ 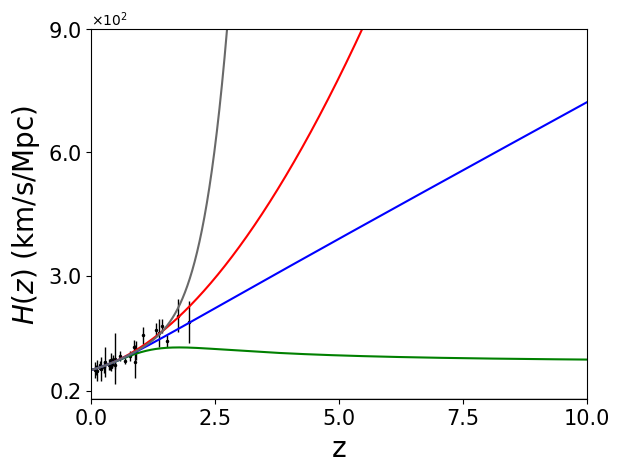}
    \caption{Luminosity distance (top panel) compared with Pantheon+ catalog \cite{Scolnic:2021amr} and the background evolution  $H(z)$ (bottom panel) compared with Cosmic Chronometers (CC) data \cite{Stern:2009ep,Moresco:2015cya,Zhang:2012mp,Moresco:2016mzx,Ratsimbazafy:2017vga, Moresco:2012jh}, considering different orders of the Pad\'e series, together with the Taylor series expansion up to fifth order in redshift. We use the \textit{fiducial} values of $\{q_0=-0.55, j_0=1, l_0=0.685, s_0=-0.35, p_0=1\}$ \cite{Capozziello:2020ctn}.}
    \label{fig:dL-Hz}
\end{figure}
This can be achieved by considering, for $f(z)$, the Taylor formula for luminosity distance  in Eq.(\ref{eq:taylor_dL}), as well as  the background evolution in  Eq.(\ref{eq:Hz}). 

The complete expressions for both $d_L(z)$ and $H(z)$ are reported in Appendix \ref{app:PXX_formulas}.  The behaviors of Pad\'e polynomials with respect to the Taylor ones are presented in Fig.\ref{fig:dL-Hz}. 
The Pantheon+ SNe catalog \cite{Scolnic:2021amr} and the Cosmic Chronometers (CC) data are also showed for comparison \cite{Stern:2009ep,Moresco:2015cya,Zhang:2012mp,Moresco:2016mzx,Ratsimbazafy:2017vga, Moresco:2012jh,Jimenez:2023flo}. 

It is possible to verify that the Taylor expansion diverges for redshift higher than $z \sim 2$ while the Pad\'e approximants are able to guarantee a more stable behavior. Here it is considered a set of \textit{fiducial} values for the cosmographic parameters, namely $q_0=-0.55, j_0=1, s_0=-0.35$, $l_0=0.685$, and $p_0=1$. 
These coefficients are obtained by comparing the theoretical expression from the $\Lambda$CDM model background: 

\begin{equation}
\label{eq:H_cosmology}
    H(z) = H_0 \sqrt{ \Omega_m (1+z)^3 + \Omega_r (1+z)^4 + 1 - \Omega_r - \Omega_m}\,,
\end{equation}
and the cosmographic expression for the Hubble parameter of Eq.(\ref{eq:Hz}), assuming $H_0=70$ km/s/Mpc, 
$\Omega_r = 5 \times 10^{-5}$ and $\Omega_m=0.3$ \cite{Capozziello:2020ctn}. Clearly these are only indicative values, and for the present purposes hereafter, we decided to investigate only up to the third cosmographic parameter (i.e., fourth derivatives order) assuming the others as $l_0=p_0=0$, without loss of generality.

\begin{figure}[b!]
    \centering
    \includegraphics[scale=0.35]{ P1_q0.png}
    \includegraphics[scale=0.35]{ 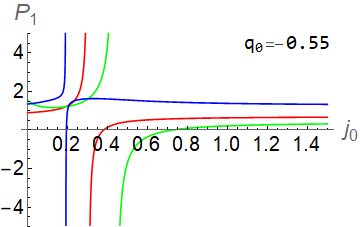}
    \includegraphics[scale=0.35]{ P1_j0barr.png}
    \includegraphics[scale=0.35]{ 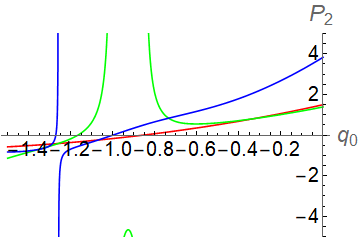}
    \includegraphics[scale=0.35]{ 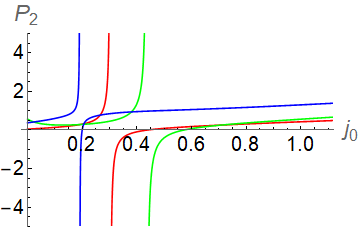}
    \includegraphics[scale=0.35]{ 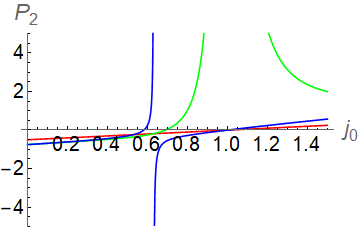}
    \includegraphics[scale=0.35]{ 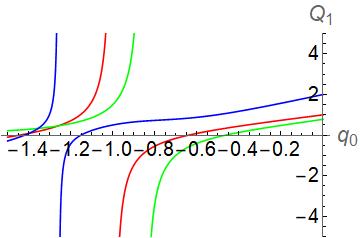}
    \includegraphics[scale=0.35]{ 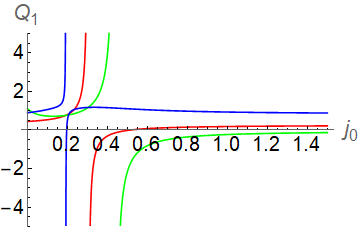}
    \includegraphics[scale=0.35]{ 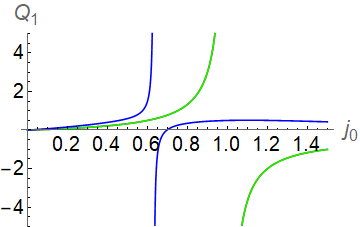}
    \includegraphics[scale=0.35]{ 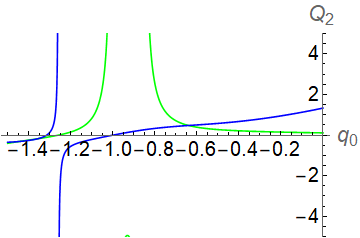}
    \includegraphics[scale=0.35]{ 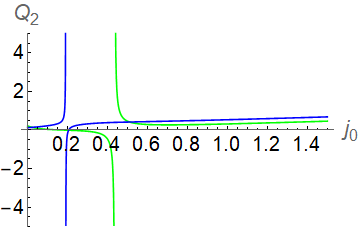}
    \includegraphics[scale=0.35]{ 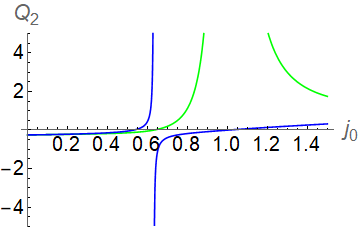}
    \includegraphics[scale=0.35]{ 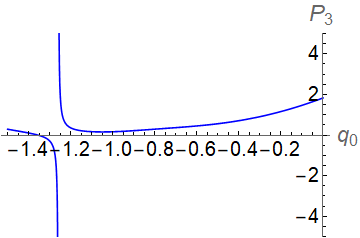}
    \includegraphics[scale=0.35]{ 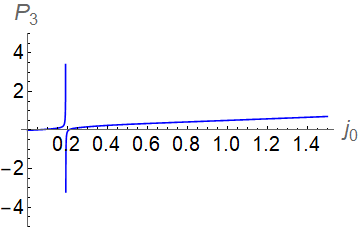}
    \includegraphics[scale=0.35]{ 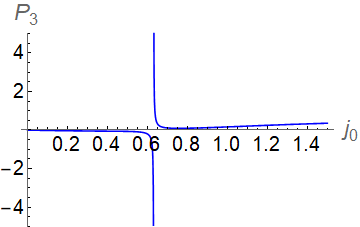}
    \caption{$P_1$, $P_2$, $P_3$ and $Q_1$, $Q_2$ coefficients behavior of Eqs. (\ref{eq:f_P21})-(\ref{eq:f_P32}) in function of $q_0$ (on the left), $j_0$ (on the center and right) for the three Pad\'e series. 
    The red curves are overlapped with the green ones in the right column for $P_1$ and $Q_1$ coefficients . 
    The values considered for the cosmography parameters on the left column are $j_0=1$, $s_0=0$, and $l_0=0$. In the central column, we fix $q_0=-0.55$, while we assume $q_0=-1$ in the right one (also considering $s_0=l_0=0$). }
    \label{fig:Pade_coeff_q0}
\end{figure}

It can be seen that by increasing the orders  in  polynomials, there is a significant difference  in both $d_L(z)$ and $H(z)$. Among the three  Pad\'e approximants, $P_{32}$ stands out for its convergent behavior. Thus, thanks to a better stability, it seems convenient  taking into account higher order polynomials, although these ones show greater complexity for  the equations of  coefficients. To better understand the role of  coefficients, it is worth taking into account   Fig.\ref{fig:Pade_coeff_q0}. Here the Pad\'e coefficients of Eqs.(\ref{eq:f_P21})-(\ref{eq:f_P32}) are reported in function of $q_0$ (left column) assuming $j_0 = 1$ (for simplicity, it is $s_0=l_0=p_0=0$), while, in the central and right column, it is showed the $j_0$ dependencies assuming $q_0=-0.55$ (central) and $q_0=-1$ (right).

\begin{figure}[t!]
    \centering
    \includegraphics[scale=0.35]{ 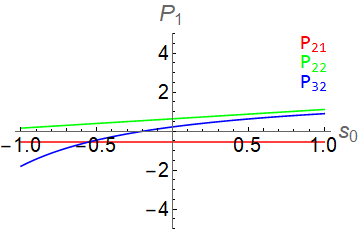}
    \includegraphics[scale=0.35]{ 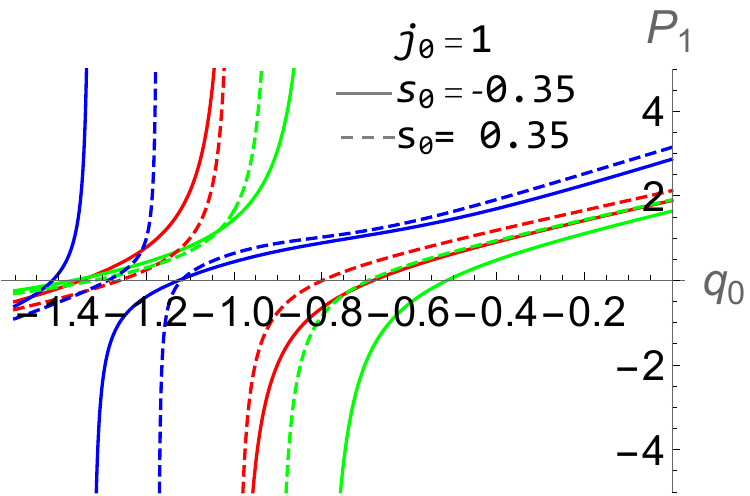}
    \includegraphics[scale=0.35]{ 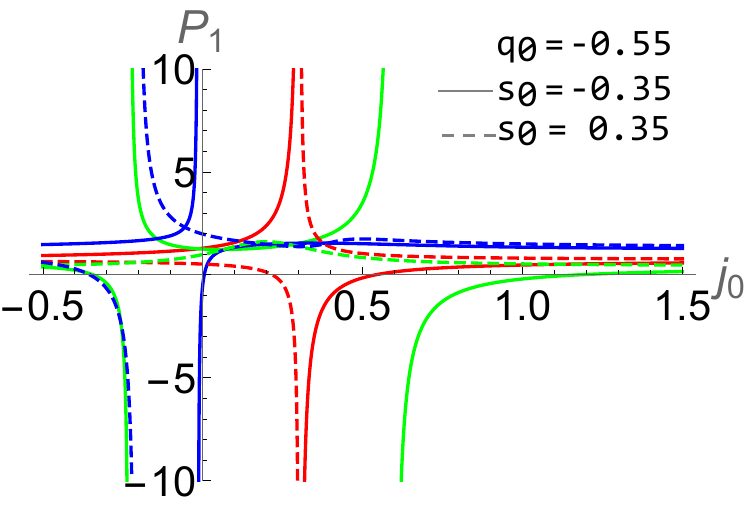}
    \includegraphics[scale=0.35]{ 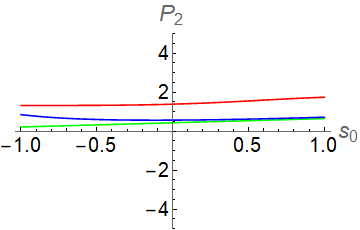}
    \includegraphics[scale=0.35]{ 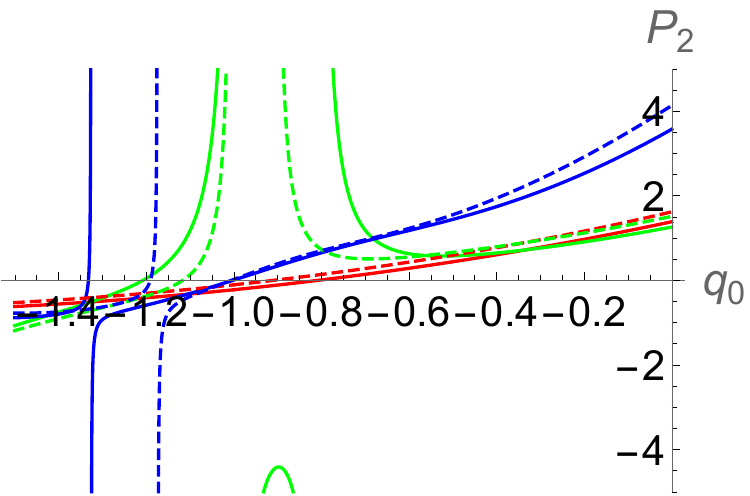}
    \includegraphics[scale=0.35]{ 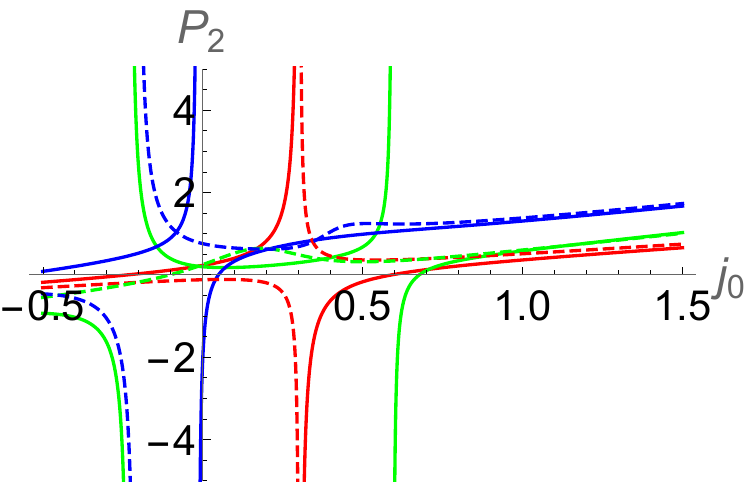}
    \includegraphics[scale=0.35]{ 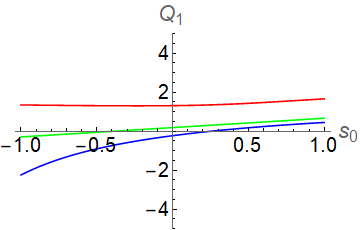}
    \includegraphics[scale=0.35]{ 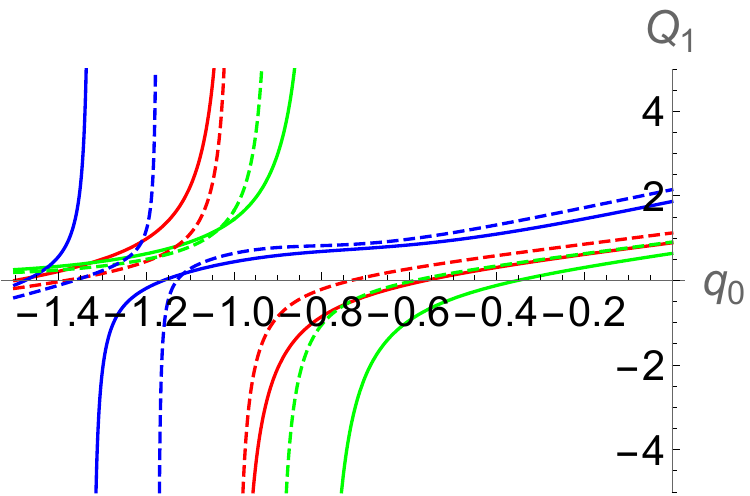}
    \includegraphics[scale=0.35]{ 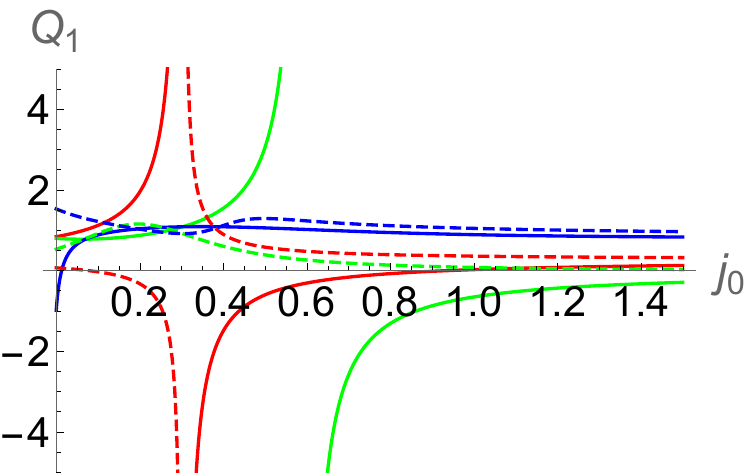}
    \includegraphics[scale=0.35]{ 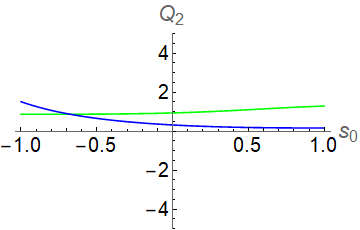}
    \includegraphics[scale=0.35]{ 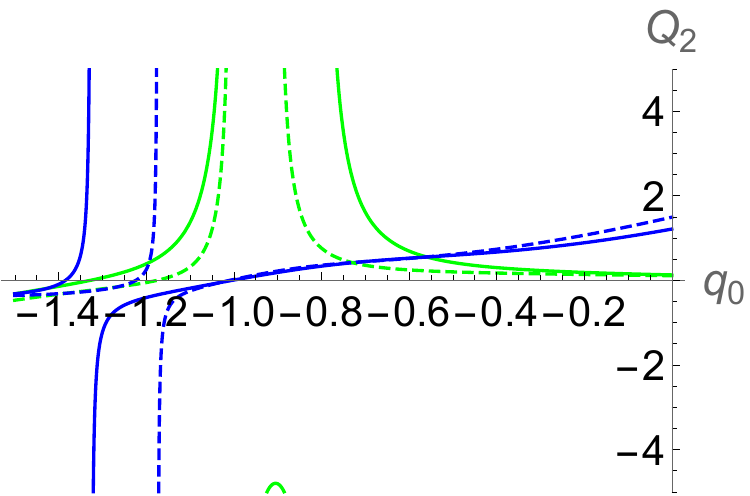}
    \includegraphics[scale=0.35]{ 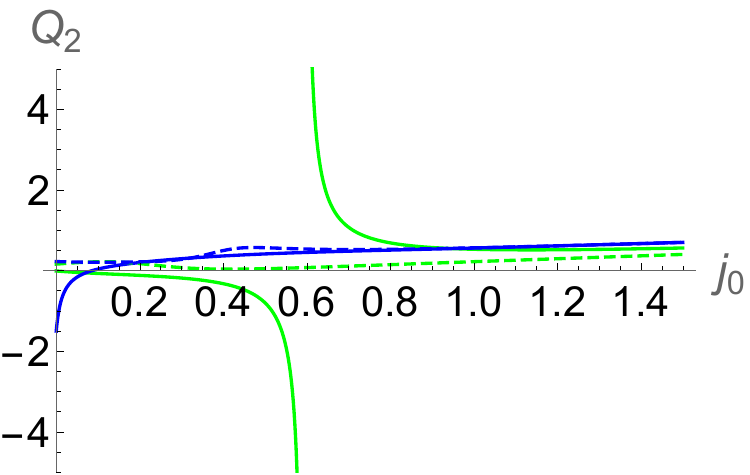}
    \includegraphics[scale=0.35]{ 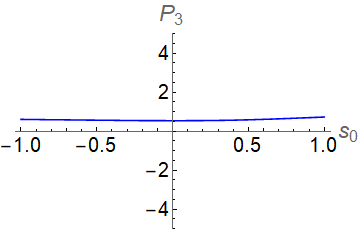}
    \includegraphics[scale=0.35]{ 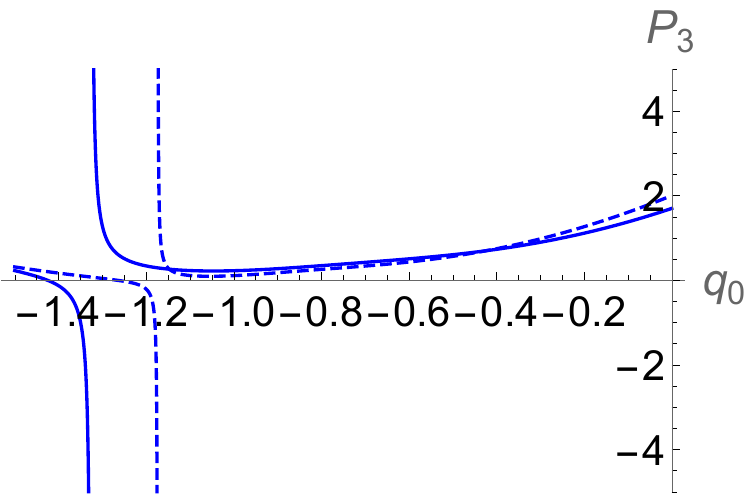}
    \includegraphics[scale=0.35]{ 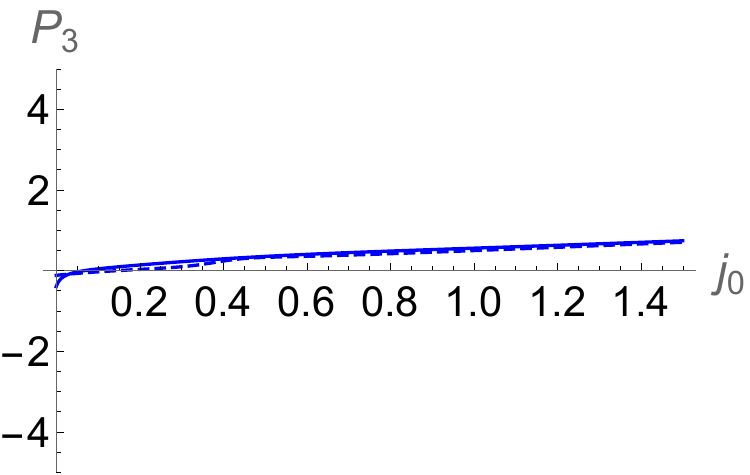}
    \caption{
    $P_1$, $P_2$, $P_3$ and $Q_1$, $Q_2$ coefficients behavior of Eqs. (\ref{eq:f_P21})-(\ref{eq:f_P32}) in function of $s_0$ (on the left), with $q_0=-0.55$ and $j_0=1$. These Pad\'e series are presented also in function of $j_0$, with $q_0=-0.55$, $s_0=-0.35$ (center) and $s_0=0.35$ (right). }
    \label{fig:Pade_coeff_s0}
\end{figure}

It is possible to state that, depending on the series analysed and the cosmographic values assumed, Pad\'e coefficients always diverge for a specific value. For example, in the case of $P_{32}$ approximation, divergence appears  at $q_0=-1.3$, which is outside the values of physical interest (i.e $q_0=-0.55$), while, for the $P_{21}$ series,  the divergence occurs at $q_0=-0.8$, which is closer to the \textit{fiducial} value. This obviously depends on the value of $j_0$ chosen. The same happens for the jerk parameter, where, depending on the value of $q_0$ set, the divergence moves to greater $j_0$ values as $q_0$ is smaller. 
The existence of this divergence then defines the range of  parameters exploration. At the same time, the Pad\'e coefficients can be directly constrained by  data, and take values approximately in the range [-2:2] \cite{Dutta:2019pio,Zhou:2016nik, Liu:2021oaj}. This additional information can be used to exclude ranges of cosmographic parameter values that generate coefficients far from observational constraints.

Also, it is worth noticing  that higher-order cosmographic parameters can play an important role for $P_{32}$ approximant. Indeed, in Fig.\ref{fig:Pade_coeff_s0}, we explore the dependence with the snap value, assuming $q_0=-0.55$, $j_0 = 1$ and $l_0=p_0=0$, while, in the central and right columns,  we show the $q_0$ and $j_0$ dependencies  assuming $s_0=-0.35$ and $s_0=0.35$.  From the left column, we note that, introducing a non-zero value for $s_0$ does not significantly change the value of Pad\'e coefficients: they show a constant behavior for any value of the snap parameter, with the exception of the series $P_{32}$, where $P_1$, $Q_1$, $Q_2$ seem  sensible to its value. On the other hand, we see that assuming a no-zero value for $s_0$, it can significantly change the divergence value of $j_0$ (right column), going so far as to eliminate it from the range of interest for the series $P_{32}$ and $P_{22}$ in the case of $s_0=0.35$ (dashed line). Thus, we can conclude that although the cosmographic parameters of higher orders to the jerk are generally not well constrained by  data \cite{Benetti:2019gmo, Capozziello:2020ctn} and can therefore be set at constant values, their values can affect the stability of  Pad\'e series coefficients, and can therefore be taken into account to avoid numerical problems in the analysis. Unfortunately, such a $s_0=0.35$ does not remove the divergence in $q_0$ (central column), that is in $q_0=-0.9$ for $P_{22}$ and $q_0=-1.2$ for $P_{32}$. 
At the same time, by exploring the range $s_0<0$, it shifts the singularity of $q_0$ to smaller values, but very slowly, i.e. for $s_0=-2$, the divergence is moved to $q_0=-1.4$. 
Let us stress that we cannot consider the value of $s_0$ significantly different from zero because these parameters have physical meanings and the observations give us indicative but important constraints on the values to explore \cite{Benetti:2019gmo}. A rigorous analysis of the divergence shows that it cannot be removed, at most, it can be shifted to values not  interesting for analysis. This point will be taken up in the Method section, where the cosmographic parameter priors are discussed.

\section{The $f(z)$CDM approach} 
\label{sec:theoryfzCDM}

Let us now move beyond the standard cosmography as introduced in the previous sections, which we refer to hereafter as \textit{vanilla}, and introduce the possibility of considering a cosmographic series to parameterize the dark energy behavior. In other words, the cosmographic series, here, is not used to parameterize the evolution of the universe in its entirety but it is placed within a cosmological model  in view of describing the dark energy contribution. This approach allows us to assume no a priori nature or evolution for the current expansion of the universe. In the following picture, the analysis of cosmographic parameters is required  to  give  information on cosmic expansion and then on dark energy. 
Furthermore, the approach allows to use  both early and late time data to constrain the entire evolution of the universe at all scales. In this way, it is not necessary to fix the values of cosmological parameters, i.e., matter density, as it is necessary to do in the standard cosmographic analyses, but it can be left as a free parameter to be restricted with the data at the same time as the deceleration and jerk values.

{ {
Let us recall the Friedmann's equation for a Friedmann-Lema\^itre-Robertson-Walker (FLRW)  metric
\begin{equation}\label{eq:1fri}
    H^2=\frac{8\pi G}{3} \sum_{i} \rho_i + \frac{\Lambda}{3} + \frac{k}{a^2},
\end{equation}
\begin{equation}\label{eq:2fri}
    2\dot H+3H^2=-\frac{8\pi G}{3} \sum_{i} p_i + \frac{\Lambda}{3}\,,
\end{equation}
where $\rho$ and $p$ are the densitity and pressure of the cosmological fluids, respectively, while $\Lambda$ is the cosmological constant and $k$ defines the spatial curvature.
The solution of the fluid continuity equation, $\dot{\rho}+3H(\rho+p)=0$, in terms of scale factor reads as 
\begin{equation}
\label{eq:continuitysolution}
    \rho(a)=\rho_0 a^{-3(1+w)}
\end{equation}
 where $w$ is the equation-of-state (EoS) of the cosmological fluid, defined as the ratio 
$w \equiv \frac{p}{\rho}$. So, we can rewrite the first Friedmann equation, i.e the background evolution of the universe of Eq.(\ref{eq:H_cosmology}), making explicit the evolutions of the cosmological fluids we are considering
\begin{equation}
H(z) = H_0 \sqrt{\Omega_k a^{-2} + \Omega_m a^{-3} + \Omega_r a^{-4} + \Omega_{\Lambda} a^{-3(1+\omega)}}\,
\end{equation}
where we define $\Omega_{i} = \frac{8 \pi G \rho_{i0}}{3H_0^2}$, with the pedix "0" indicating the present time, and $\omega \equiv w_{DE}$ the EoS of the dark energy. }}

{ {Given the particular choice of $\omega$ then it is possible to consider different natures of dark energy, such as constant and equal to $-1$, as considered in the $\Lambda$CDM model, or constant but different from $-1$, as in the $\omega$CDM model. Or linearly scale-dependent as the CPL model, $\omega(a)=\omega_0+\omega_a (1-a)$ \cite{CHEVALLIER_2001,2003PhRvL..90i1301L}, or with other scale-dependence as the JBP model \citep{2005MNRAS.356L..11J}, the exponential \cite{2019PhRvD..99d3543Y} or BA model \cite{2008PhLB..666..415B}. We refer the reader to some interesting reviews \cite{Brax:2017idh,Arun:2017uaw,Tawfik:2019dda,Frusciante:2019xia,Poulin:2023lkg}. }}

{ {The effective dark energy EoS, $\omega$, can then be written as a function of the scale factor, and can generically be denoted as $f(a)$ remaining consistent with the Friedmann and continuity equations used to build the cosmological equations. }}

{ {At the same time, the solution of the continuity equation, Eq.(\ref{eq:continuitysolution}),  can be approximated at low redshift as a Taylor expansion \cite{Dutta:2019pio}
\begin{equation}
    \rho_{DE}(a) = \rho_0 + \rho_1(1 - a) + \rho_2(1 - a)^2 + ..
\end{equation}
which, rewritten in terms of redshift, can take the form of a Pad\'e expansion around $z=0$, i.e as a $P_{22}(z)$ of Eq.(\ref{eq:f_P22}):
\begin{equation}
    \rho_{DE}(z) = \frac{\rho_0 + (2\rho_0 + \rho_1)z + (\rho_2 + \rho_1 + \rho_0)z^2}{1 + 2z + z^2} \equiv P_{22}(z)
\end{equation}
}}

{ {Then, we can consider the dark energy density as a general Pad\'e expansion, $P_{nm}(z)$, and assume a background evolution equation as \cite{Benetti:2019gmo}
}}

\begin{equation}
\label{eq:H_full}
H(z) = H_0 \sqrt{\Omega_k (1+z)^2 + \Omega_m (1+z)^3 + \Omega_r (1+z)^4 + \Omega_f P_{nm}(z)}\,,
\end{equation}

with $\Omega_k + \Omega_m + \Omega_r + \Omega_f = 1$ and $P_{nm}$(z) the chosen Pad\'e series, as introduced in the previous section. Alternatively, one can avoid to take a specific form and reconstruct the function with the data, as shown in \cite{Dutta:2018vmq}.

\begin{figure}[!]
    \centering
    \includegraphics[scale=0.4]{ 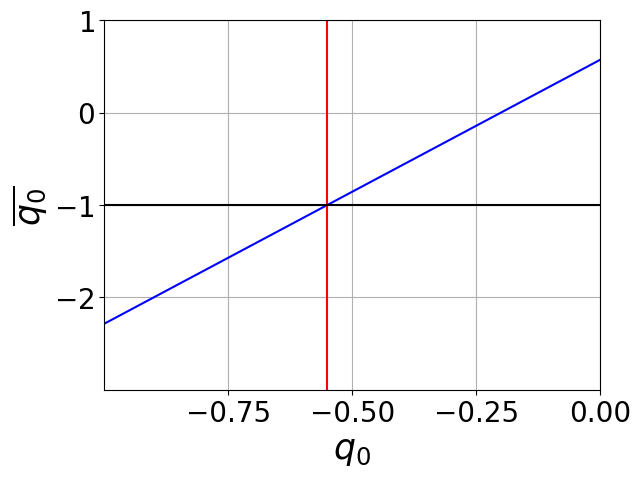}
    \includegraphics[scale=0.4]{ 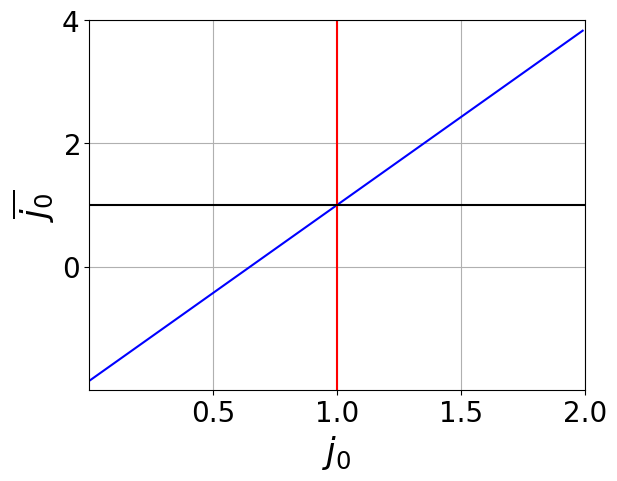}
     \caption{The $\overline{q_{0}}$ and $\overline{j_{0}}$ parameters from $f(z)$CDM model as functions of the $q_0$ and $j_0$ cosmographic parameters, respectively, as described by Eqs.(\ref{eq:q0t})-(\ref{eq:j0t}) . Red line indicates the cosmographic \textit{fiducial} values $q_0=-0.55$ and $j_0=1$, while the black lines stand for the corresponding $f(z)$CDM parameters, $\overline{q_{0}}=-1.00$ and $\overline{j_{0}}=1.00$.}
    \label{fig:relationq0_barq0}
\end{figure}

While in the \textit{vanilla} cosmography the whole background evolution is given by the Pad\'e approximation of Eqs.(\ref{eq:f_P21})-(\ref{eq:f_P32}), in Eq.(\ref{eq:H_full}),  $H(z)$ is given by the evolution of the cosmological fluids and only the current universe expansion is parameterised by the Pad\'e series, weighted by the density $\Omega_f$. This makes it possible to capture the behavior of interest, eliminating the degeneracy between cosmographic parameters and the matter density, which here evolves  as predicted by the standard cosmological model. 
Thus, the cosmographic parameters of the $f(z)$CDM approach are different from those of \textit{vanilla} cosmography. The relation between the \textit{vanilla} $q_0$ and $j_0$, and those constrained by Eq.(\ref{eq:H_full}) (hereafter  $\overline{q_0}$, $\overline{j_0}$, etc.) can be determined by equating the derivatives of the two expressions, i.e $H(z)$ and the considered Pad\'e serie $P_{nm}$(z), at $z=0$:
\begin{equation*}
    \frac{H'(0)}{H_0} = P'_{nm}(0)
\end{equation*}
\begin{equation*}
    \frac{H''(0)}{H_0} = P''_{nm}(0)
\end{equation*}
\begin{equation*}
\vdots
\end{equation*}
\begin{equation}
    \frac{H^{(n+m)}(0)}{H_0} = P_{nm}^{(n+m)}(0)
    \label{eq:nmorder}
\end{equation}
so that, from the prime derivatives, it is possible to determine $\overline{q_0}$ as a function of $q_0$, from the second derivatives it is possible to determine  $\overline{j_0}$ as a function of $j_0$
and the further orders from the higher derivatives. Specifically, for the first two orders, it is:
\begin{equation}
    \overline{q_{0}} = \frac{2 \Omega_m-1 - 2 q_0}{\Omega_m-1}
    \label{eq:q0t}
\end{equation}
\begin{equation}
    \overline{j_{0}} = \frac{3 + 4 q_0^2 + q_0 (8 - 12 \Omega_m) - 2 j_0 (\Omega_m-1) - 12 \Omega_m + 10 \Omega_m^2}{(\Omega_m-1)^2}
    \label{eq:j0t}
\end{equation}

where we can set $\Omega_r$ = $\Omega_k$ = 0. 
These relations are plotted in Fig. \ref{fig:relationq0_barq0}, where it is left explicit that the \textit{fiducial} values $q_0=-0.55$, $j_0=1$
(red line) correspond to $\overline{q_0}=-1.0$, $\overline{j_0}=1.0$ 
(black line). 

\section{Method and Results}
\label{sec:Method}

Let us illustrate now  the method used to estimate the parameters of  $f(z)$CDM model, as well as the codes adopted in this analysis, and the observational data set employed.  As mentioned earlier, the $f(z)$CDM approach, unlike \textit{vanilla} cosmography, allows us to use a joint data set of large- and small-scale data. Thus, a combined information from the most robust data releases of several independent measurements is selected, covering from early to late time:
\begin{itemize}

\item Cosmic Microwave Background Radiation (CMBR) measurements, through the Planck (2018) data \cite{Planck:2019nip}, considering both temperature power spectra (TT) over the range $\ell \in [2 - 2508]$ and low-$\ell$ (2 - 29) temperature-polarization cross-correlation likelihood, and HFI polarization EE likelihood at $\ell \leq 29$. It is also considered CMB lensing reconstruction power spectrum ~\cite{Planck:2019nip,Planck:2018lbu};

\item Baryon Acoustic Oscillation (BAO) distance measurements, using 6dFGS ~\cite{Beutler:2011hx}, SDSS-MGS~\cite{Ross:2014qpa}, and 
BOSS DR12~\cite{Alam:2016hwk} surveys;

\item Supernovae Type Ia (Pantheon+ sample), that is the latest compilation of 1701 SNIa covering the redshift range $[0.01 : 2.3]$~\cite{Scolnic:2021amr,Brout:2022vxf}, calibrated with SH0ES \cite{Riess:2021jrx} \footnote{The sample is avalaible at http://PantheonPlusSH0ES.github.io, with also the Statistical and Stat+Systematic Covariance matrices.} using Chepeid information by Hubble Space Telescope (HST). This sample has a more accurate low redshift data set with respect to the Pantheon catalog and applies more rigid selections on supernova light curves.

\item Cosmic Chronometers (CC) data, measurements of the expansion rate $H(z)$ from the relative ages of passively evolving galaxies \cite{Jimenez:2001gg,Simon:2004tf,Stern:2009ep,Moresco:2015cya,Zhang:2012mp,Moresco:2016mzx,Ratsimbazafy:2017vga, Moresco:2012jh}.
\end{itemize}

The modifications have been implemented according to Eq.(\ref{eq:H_full}), with Pad\'e approximant given by Eq. (\ref{eq:Hz_P21})-(\ref{eq:Hz_P32}), in the Boltzmann solver \textit{Code for Anisotropies in the Microwave Background} ({\sc CAMB})\cite{Lewis:1999bs}. The Monte Carlo Markov Chain method of the parameter estimation packages {\sc CosmoMC}~\cite{Lewis:2002ah} has been also used to constrain the parameters and perform the statistical analysis. 
 Furthermore, the standard set of cosmological parameters are considered free in our analysis: the baryon density ($\Omega_bh^2$), the cold dark matter density ($\Omega_ch^2$), the ratio between the sound horizon and the angular diameter distance at decoupling ($\theta$), the optical depth ($\tau$), the primordial scalar amplitude ($A_s$), and the primordial spectral index ($n_s$). To parameterize the current universe evolution, $\overline{q_0}$ and $\overline{j_0}$ are also constrained. Note that these parameters are correlated with the cosmographic deceleration and jerk parameters by Eq.(\ref{eq:q0t})-(\ref{eq:j0t}), with the central fiducial value of $\overline{q_0}=-1$ and $\overline{j_0}=1$. Very large and flat priors for each parameter are considered. In particular, $\overline{q_0}$ and $\overline{j_0}$ are free to vary into the ranges [-3 : 0.5] and [-2 : 4], respectively, that corresponds to cosmographic values of ${q_0} \in [-1.25:-0.025]$ and ${j_0} \in [-0.05 : 2.05]$ (see Fig.\ref{fig:relationq0_barq0}).
 
  For such prior ranges, a reasonable stability in Pad\'e's coefficients for the $P_{32}$ approximation was found, assuming $\overline{s_0}=1.2$. Indeed, fixing such a value, it is possible to remove the divergences shown in Fig.\ref{fig:Pade_coeff_q0}-\ref{fig:Pade_coeff_s0}. 
 
Unfortunately, for the other two considered Pad\'e parameterizations, it was not possible to improve stability using values of $\overline{s_0}$ other than zero so, for $P_{21}$ and $P_{22}$, $\overline{s_0}=0$ in the analysis, as constrained in previous results \cite{Benetti:2019gmo}. This, therefore, marks a point in favor of $P_{32}$, whose complexity, compared to other parameterizations,  allows mathematical divergences to be shifted to a range outside the interest of our analysis. 

It is relevant to stress that, as shown in the left-hand column of Fig.\ref{fig:Pade_coeff_s0}, Pad\'e's coefficients take on similar values for each value of $s_0$ considered, and the only task of the snap parameter is to shift the divergence of $q_0$ and $j_0$ for values away from those of interest. It has been shown in previous works that it is not possible to constrain this parameter tightly, and that large uncertainties are related to its value \cite{Capozziello:2020ctn, Capozziello:2019cav}. The negligible impact of the snap parameter on the observational predictions can also be seen by analysing the temperature spectra of the CMBR as $\overline{s_0}$ varies, which are fully superimposed. In addition, it was tested whether the variation of $\overline{s_0}$ changes the parameter constraints, finding that the value of $\overline{s_0}$ does not affect the value of the cosmological parameters and has the only role of varying  the range width of cosmographic values, but not the value of the posterior peak of both $\overline{q_0}$ and $\overline{j_0}$. In the case of a not optimal choice of $\overline{s_0}$, the confident levels regions of the analysis are cut and incomplete, still leaving the preferred value of the parameters in 1$\sigma$ agreement.  

\begin{table*}
\caption{Mean values and 1$\sigma$ uncertainties for the cosmological and cosmographic parameters of each model considered in this work. The data set used join information of CMB TTTEEE+lowE+lensing, BAO, Pantheon+ and CC. In the upper section of the table, some of the most significant primary parameters of our analysis, in the middle part some of the derived parameters of interest  {, in the lower part the best-fit $\Delta_{\chi_\text{eff}^2} = 
\chi_\text{eff}^2 (\text{P$_{nm}$}) - 
\chi_\text{eff}^2 (\text{P$_{21}$})$ and the $\Delta \text{DIC} = 
\text{DIC} (\text{P$_{nm}$}) - 
\text{DIC} (\text{P$_{21}$})$. A negative $\Delta$ value means that the Pad\'e model is supported by data over the reference $P_{21}$ one.}}
\label{tab:analysis}
\centering 
\scalebox{1.1}{
\begin{tabular}{ c| c| c| c } 
\hline\hline
 & $P_{21}$ &$P_{22}$ &$P_{32}$ \\
 \hline
 100$\Omega_b h^2$ & $2.240 \pm 0.014 $& $2.244 \pm 0.016 $ & $2.240 \pm 0.014 $\\
 $\Omega_c h^2$ & $ 0.1195 \pm 0.0010$ & $ 0.1187 \pm 0.0011$ & $ 0.1196 \pm 0.0012$\\
  $n_s$ &  $0.9661 \pm 0.0038$&  $0.9673 \pm 0.0041$& $0.9656 \pm 0.0045$\\
  $\overline{q_0}$ &  $-0.86 \pm 0.06$ &  $-1.06 \pm 0.12$ &$-0.70 \pm 0.09$\\
 $\overline{j_0}$ & $0.45 \pm 0.17$ & $1.71 \pm 0.38$ & $0.32 \pm 0.24$\\
 \hline
 $H_0$ &  $67.00 \pm 0.76$ & $66.51 \pm 0.84$ & $66.45 \pm 0.70$ \\
 $\sigma_8$ & $0.8065 \pm 0.0087$ &  $0.7980 \pm 0.0084$ & $0.8035 \pm 0.0100$\\
 $\Omega_m$ & $0.3177 \pm 0.0073$& $0.3207 \pm 0.0086$& $0.3233 \pm 0.0069$\\
\hline
 $\Delta_{\chi_\text{eff}^2}$ & $ 0 $& $ + 3.1 $& $-0.9$\\
  $\Delta \text{DIC}$ & $ 0 $& $ -1.9 $ & $-2.5$\\
  \hline
   \hline
\end{tabular}}
\end{table*}

\hfill

\begin{figure}
    \centering
    \includegraphics[scale=0.45]{ 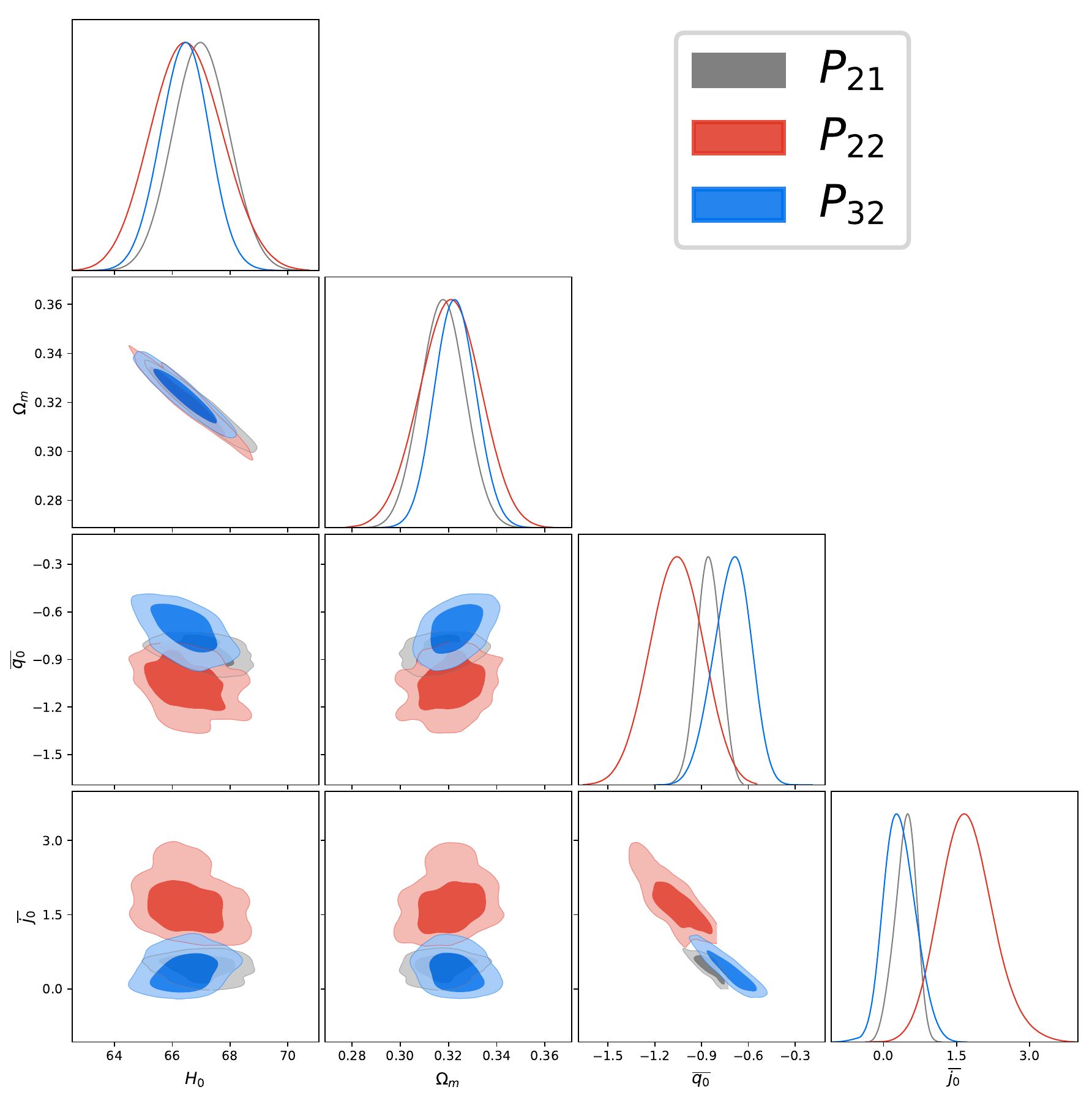}
    \caption{2D Confidence Levels and posterior distributions for the cosmological and cosmographic parameters from $f(z)$CDM model with background evolution of Eq.(\ref{eq:H_full}), considering $f(z)$ given by $P_{32}$, $P_{22}$, $P_{21}$ Pad\'e approximations.} 
    \label{fig:tri_analysis}
\end{figure}

In Tab.\ref{tab:analysis} and Fig.\ref{fig:tri_analysis},  the results of our analysis are presented.   
Firstly, we note that the $P_{21}$ and $P_{32}$ show a similar behavior, constraining $\overline{q_0}$ higher then $-1$ and $\overline{j_0}$ lower then unit, while $P_{22}$ prefers more negative deceleration parameter values and higher jerk. 
This is fully compatible with previous constraints of $P_{22}$ using CMBR data in  Ref.\cite{Benetti:2019gmo}, where the main difference between that analysis and the present one is in the use of a $H_0$ prior, given by the Riess et al. (2022) \cite{Riess:2019cxk}, i.e. $H_0 = 74.03 \pm 1.42$ km/s/Mpc. This forces results in \cite{Benetti:2019gmo} to be constrained   at a more negative deceleration parameter, i.e. $\overline{q_0}= -1.19 \pm 0.10$, which is, in any case, in agreement in 1$\sigma$ with present results. 

It is worth noticing that the best fit values for $\overline{q_0}$ and $\overline{j_0}$ of the present analysis corresponds to cosmographic ${q_0}$ and $j_0$ values as showed in Tab.\ref{tab:cosmographic}. This allows to compare these constraints with previous cosmographic analysis performed by using large scale data, i.e. without including CMBR information. So, although the present data set includes both small and large scale measurements, the results are fully compatible with Ref.\cite{Capozziello:2020ctn}, and agree in 2$\sigma$ with Ref.\cite{Capozziello:2018jya, DAgostino:2023cgx}. 
As expected, the inclusion of CMBR data gives more control over high-order terms, better constraining the term $j_0$ and limiting it to values below unity, whereas, using only large-scale data does not give a good bound over this term, which is constrained to higher values with higher uncertainties. This is also true for higher orders, which are often associated with an error higher than their mean value or are not bound at all \cite{DAgostino:2023cgx}. Trying to leave these parameters unconstrained, in MCMC analysis, not only does not bring any interesting constraints but also risks the oversample of the model.

 {
In order to better discuss the performance of the three Pad\'e series we are analyzing, we can consider some useful tools widely used in the literature, i.e the bayesian criteria AIC (Akaike information
criterion), BIC (Bayesian information criterion) and DIC (Deviance information criterion) \cite{Rezaei:2021qpq,Spiegelhalter:2002yvw,Akaike:1974,Schwarz:1978,Spiegelhalter:2002yvw,Kunz:2006mc,Trotta:2008qt}
\begin{equation}
\begin{split}
\text{AIC}:=& \chi_\text{eff}^2 + 2 N, \\
\text{BIC}:=& \chi_\text{eff}^2 + N \ln(k), \\
\text{DIC}:=& \chi_\text{eff}^2 + 2 p_\text{D}, \\
\end{split}
\end{equation}
where $\chi_\text{eff}^2$ is the effective $\chi^2$ corresponding to the maximum likelihood (i.e. the best fit), $N$ is the number of free parameters of the analysis, $k$ is the number of data points used for the constraints and $p_\text{D} = \overline{\chi}_\text{eff}^2 - \chi_\text{eff}^2$, with the bar indicating the average of the posterior distribution. 
In these criteria, the first term accounts for the fitting quality of the model while the second one represents the model complexity. 
While AIC is a simple indication, the BIC criteria starts to show some Bayesian information, and  DIC accounts for the better representation of the Bayesian complexity of the model, taking into account its average performance.}

 {The strategy is then to calculate the difference of these AIC, BIC, and DIC values with the reference one, and evaluate with a Jeffreys' scale their goodness, where $\Delta>10$, $>5$, $>2$ provide, respectively, strong/moderate/weak evidence against the considered Pad\'e model. Instead, a negative value of $\Delta$ values means the analysed Pad\'e model is supported by data over the reference one. }

 {In our case, the total number of free parameter of the analysis, $N$, is the sum of the $6$ free parameter of the standard cosmological model with 2 cosmographic parameters ($\overline{q_0}, \overline{j_0})$ and eventually more $21$ Planck nuisance parameters \cite{Planck:2019nip} used by the likelihood marginalization. This value is fixed for the three models considered. In fact, although $P_{32}$ uses more padè coefficients than $P_{21}$, these are written and analyzed in terms of the deceleration and jerk parameters, as described in the previous section an detailed in \ref{app:PXX_formulas}. 
Regarding the value of $k$, the number of data point used in BIC criteria, this is also the same for all the analyses conducted here, where we consider only one dataset. 
Then both AIC and BIC criteria give the same result, which is the difference in the $\chi_\text{eff}^2$ of the analysis.  We choose to consider the simpler $P_{21}$ as the reference model and report the 
$\Delta_{\chi_\text{eff}^2} = 
\chi_\text{eff}^2 (\text{P$_{nm}$}) - 
\chi_\text{eff}^2 (\text{P$_{21}$})$ 
values in the last lines of the Tab.~\ref{tab:analysis}. 
At the same time, we also show the $\Delta$DIC value. 
Based on Jeffrey's scale, there is no significant data preference shown against any of the three analyses, although $P_{32}$ is weakly preferred over $P_{21}$ for the chosen dataset following the DIC criteria and $P_{22}$.  In general, we see that the improvement of complexity seems to be favored by the data, although not decisively for the choice of Pad\'e approximant.}

\begin{figure}
    \centering
        \includegraphics[scale=0.45]{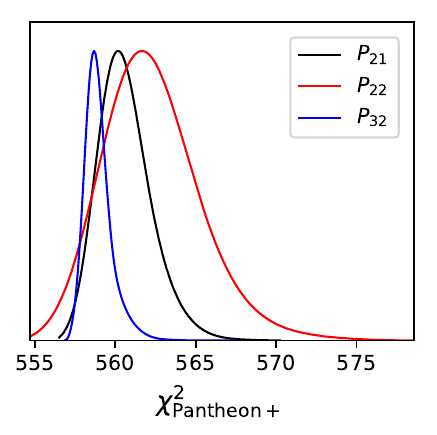}
            \includegraphics[scale=0.45]{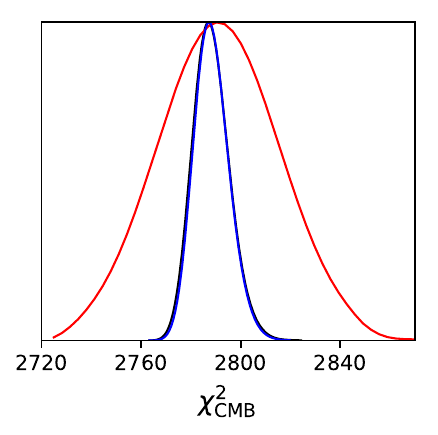}
            \includegraphics[scale=0.45]{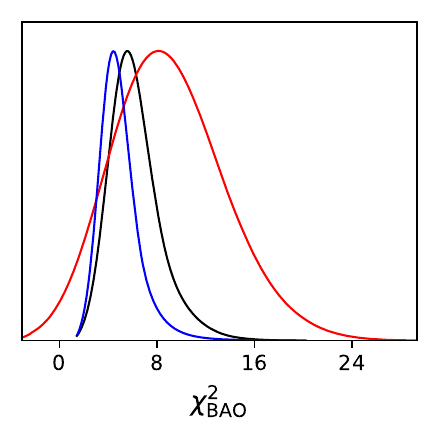}
                \includegraphics[scale=0.45]{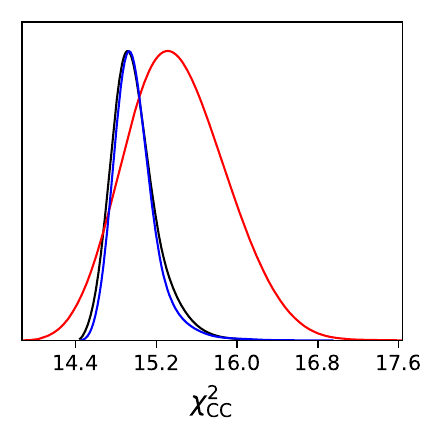}
    \caption{ $\chi^2_{eff}$ posterior distributions for the $f(z)$CDM model analysis using the data discussed in Sec.4.} 
    \label{fig:chi2}
\end{figure}

 {Finally, we also analyze in Fig.\ref{fig:chi2} the contribution of each data separately to the $\chi_\text{eff}^2$ value to better understand the weight of each on the final results. 
Looking at the posterior distributions, we see that the analysis with the $P_{22}$ model has a wider error-bar while the approximants with one-order difference between numerator and denominator succeed in being more piked. At the same time, we can appreciate that the largest difference in the value of $\chi_\text{eff}^2$ is given by the SNe dataset Pantheon+, where the value is lowered for $P_{31}$ better than for $P_{21}$, and which in general has a slightly larger dispersion for the latter. }
}

\section{Discussion and Conclusions} 
\label{sec:Conslusions}

In this paper, cosmography is used to describe the current acceleration of the universe beyond the  $\Lambda$CDM model.  Specifically, we consider the  $f(z)$CDM model, where radiation and matter evolution are described as in the standard cosmological model and the current observed expansion is parameterised with a Pad\'e cosmographic series, as Eq.(\ref{eq:H_full}),without assuming any specific dark energy component. This allows to constrain the model directly by data at all scales, i.e. both CMBR and late time data. The combination of these data sets  allows to simultaneously constrain the cosmographic parameters with late time data,   matter density (usually fixed in the \textit{vanilla} cosmography), and  CMBR information. It is important to emphasize that our  Pad\'e cosmography parameterization only describes the behavior of dark energy, i.e. the current universe expansion, leaving  matter to evolve in its standard description. Instead, the \textit{vanilla} cosmography takes into account the whole current evolution, given by  dark energy and  (ordinary and dark) matter. The present values of cosmographic parameters can be traced back to the \textit{vanilla} ones using Eqs. (\ref{eq:q0t})-(\ref{eq:j0t}) and the analysis shows that, using both late and early times data, it is possible to constrain cosmographic parameters up to the third order with significant accuracy. In this context, the CMBR can operate a significant control on the constraint of the jerk parameter, improving the accuracy with respect to the error bar using only late time measurements \cite{Capozziello:2020ctn, Capozziello:2018jya, DAgostino:2023cgx}. 

The goal of this work is providing new updated constraints on cosmographic parameters by three distinct Pad\'e parameterizations, where late and early times data are used in the $f(z)$CDM context. We are also interested in  answering the  crucial issue on what  the best choice of Pad\'e series is in order to fit  cosmological and astrophysical data. Indeed, cosmography is a model-independent tool but it must still assume some parameterization. A common problem is  to fix the impact of the adopted parameterization, and derive  the optimal choice  by combining  data fitting with a minimum number of parameters. 

Here three different Pad\'e's parameterizations have been  analysed, namely $P_{32}$, $P_{22}$ and $P_{21}$. The former and the latter are the most quoted in the literature, as the difference in order between the numerator and denominator improves the convergence of the series at high redshifts. On the other hand, the second is less performing but widely used in  literature \cite{Benetti:2019gmo, Capozziello:2019cav, Dutta:2018vmq, Dutta:2019pio, Capozziello:2020ctn, Capozziello:2018jya}. We   considered it for the sake of  completeness of the analysis in view of   giving a definitive answer to the  question: \textit{Which  is the best choice for Pad\'e series?}

\begin{table}
\caption{Best fit values for the free parameters $\overline{q_0}$ and $\overline{j_0}$ for each Pad\'e approximantion considered in this work and the corresponding cosmographic values, calculated using the relations Eqs.(\ref{eq:q0t})-(\ref{eq:j0t}). }
\label{tab:cosmographic}
\centering 
\scalebox{1.1}{
\begin{tabular}{ c| c| c| c } 
\hline\hline
 & $P_{21}$ &$P_{22}$ &$P_{32}$ \\
 \hline
  $\overline{q_0}$ &  $-0.79$  &  $-1.03$  &$-0.69 $\\
 $\overline{j_0}$ & $0.22$  & $1.59$  & $0.28 $\\
 $\Omega_m$ & $0.3126$  & $0.3288$  & $0.3159 $\\
 \hline
\hline
  ${q_0}$ &  $-0.46$  &  $-0.52$  &$-0.42 $\\
 ${j_0}$ & $0.73$  & $1.19$  & $0.75 $\\
 \hline\hline
\end{tabular}}
\end{table}

It can be seen that  constraints on the cosmographic parameters of the $P_{22}$ series show some differences with  the other two choices (see Tab.\ref{tab:analysis} and \ref{tab:cosmographic}), preferring a greater $j_0$. Instead, it is noted  a similar behavior between $P_{21}$ and $P_{32}$, as expected \cite{Capozziello:2020ctn}. 
Therefore, given the similar results between the latter choices, it is necessary to consider  pros and cons of the two series to answer our question. 
On the one hand $P_{21}$ is quite simple, converges for high redshifts and shows similar results even using higher orders, such as $P_{32}$. On the other hand, it shows singularities in the definition of the Pad\'e coefficient that cannot be eliminated (see Fig.\ref{fig:Pade_coeff_q0}-\ref{fig:Pade_coeff_s0}). Instead, the greater complexity of the $P_{32}$ series manages to stem the problem by shifting the  values of $s_0$, which can  diverge, to values not interesting for the analysis. This means  larger explored parameter region and greater rigorous MCMC analysis. 

For this reason, the conclusion is that cosmography is a useful tool, which must be  properly used  by choosing the series  according to what one is interested to analyse. The choice of the specific order  should depend on the redshift range one want to explore and the number of cosmographic parameters one want to constrain.  
For late time analyses, one can consider  $P_{21}$, taking into account that the divergences of the parameters are outside the range of the values of interest. Indeed, the series has non-removable divergences in $q_0$, as shown in the left column of Fig.\ref{fig:Pade_coeff_q0}, and in $j_0$, where the value of divergence depends on the value of $q_0$. Specifically, it is seen that $P_{21}$ shows divergence for $q_0=-1$, and for $j_0<0.2$ when considering the value of $q_0=-0.55$. The values do not fall within the range preferred by the astrophysical data. On the other hand, for the exploration of high redshifts and for the use of the $f(z)$CDM approach, the $P_{32}$ series  proves to be more performing as, for appropriate values of the snap parameter, it is able to move the divergences to values even further away from those of interest, guaranteeing stability to the analysis and reliability of the results. 
Our answers are therefore only partial, and concerning the Pad\'e cosmographic series. We aim to explore, in a forthcoming paper, other rational series that seem promising, such as Chebyshev's approximants \cite{Capozziello:2019cav, Capozziello:2017nbu}, with the ultimate goal of identifying a series that can completely remove the divergences in cosmographic coefficients and allow more flexibility to fit the data.  
{
This will allow for a more rigorous analysis of the observables, using model-independent data such as the BAO's 2-point angular correlation function measurements ~\cite{BOSS:2016lsx,Carvalho:2015ica, Alcaniz:2016ryy,Carvalho:2017tuu,deCarvalho:2020ftb,Abdalla:2022yfr}, or dealing with different calibrations so that the impact of both the assumptions on the acoustic horizon, $r_d$, and on the fiducial model can be assessed. Calibrating the SNeIa of the Pantheon+ catalog with sources other than Cepheids could also be of great interest \cite{Verde:2023lmm,Kenworthy:2022jdh,Freedman:2020dne,Sandoval-Orozco:2023pit}, and the $f(z)$CDM model could prove sufficiently flexible to highlight any features of the observations.}

\section*{Acknowledgments}

We acknowledges Prof. Edivaldo Moura Santos and Dr. Rocco D'Agostino for usefull discussions.  {We also thank the anonymous referees for their suggestions for improvement.}
A.T.P. thanks for Fundação de Amparo à Pesquisa do Estado de São Paulo (FAPESP), process of number 2022/14565-9, for the financial support. M.B. and S.C. acknowledge Istituto Nazionale di Fisica Nucleare (INFN), sezione di Napoli, iniziative specifiche QGSKY and MOONLIGHT-2. We also acknowledge the use of CosmoMC package. This work was developed thanks to the National Observatory (ON) computational support.  This paper is based upon work from the COST Action CA21136, addressing observational tensions in cosmology with systematics and fundamental physics
(CosmoVerse) supported by COST (European Cooperation in Science and Technology).

\appendix

\section{Cosmographic expansions}
\label{app:PXX_formulas}

We expand here the formulas of luminosity distance and background evolution for the three Pad\'e series plotted in Fig.\ref{fig:dL-Hz}, nominally $P_{21}$, $P_{22}$ and $P_{32}$.

\begin{itemize}
  
       \item $P_{21}$ equations of the luminosity distance and the Hubble rate:

\begin{equation}
    d_L^{P_{21}}(z) = \frac{cz}{H_0}\frac{6(q_0-1)+z[-5-2j_0+q_0(8+3q_0)]}{-2(3+z+j_0z)+2q_0(3+z+3zq_0)}
\end{equation}

\begin{equation}
    \begin{split}
H^{P_{21}}(z) & = H_0(3 (j_0^2-q_0^4)+2 q_0 s_0 )z^2 +(-12 q_0^3+2 s_0) z(1+z)-6 q_0^2 (1+z)^2+ 2 j_0 (3+\\
& (6 +7 q_0)z+(3+7 q_0+q_0^2)z^2)/(-6 q_0^3 z+2 s_0 z-6 q_0^2 (1+z)+j_0 (6+(6+8 q_0)z))
    \end{split}
    \label{eq:Hz_P21}
\end{equation}

           \item $P_{22}$ equations of the luminosity distance and the Hubble rate:

\begin{equation}
    \begin{split}
    d_L^{P_{22}(z)} & = \frac{cz}{H_0}{6[10+9z-6q_0+q_0^2 z+s_0z-2q_0^2(3+7z)-q_0(16+19z)+j_0(4+(9+6q_0)z)]} \\ 
    &\{60+24z+ 6s_0z-2z^2+4j_0^2z^2-9q_0^4z^2-3s_0z^2+6q_0^3z(-9+4z) \\ 
    & +q_0^2(-36-114z+19z^2) + j_0[24+6(7+8q_0)z + z^2(-7-23q_0+6q_0^2)] \\
    &+ q_0[-96-36z+z^2(4+3s_0)]\}^{-1}
    \end{split}
\end{equation}

 \begin{equation}
    \begin{split}
        H^{P_{22}}(z) & = H_0 (24 s_0+6l_0z+72s_0z+30j_0^3z^2+6l_0z^2-45q_0^6z^2+48s_0z^2+4s_0^2z^2\\
        &-90q_0^5z(1+2z) -18q_0^4(2+21z+21z^2)+j_0^2(36+12(4+5q_0)z \\ &+(48+120q_0-5q_0^2)z^2)+ 3q_0^3(-48-144z+(-96+5s_0)z^2)\\
        &+9q_0^2(-8+2(-8+3s_0)z +(-8+l_0+12s_0)z^2)+6q_0(l_0z(1+2z)\\ &+s_0(4+23z+23z^2))+j_0(60q_0^4z^2+90q_0^3z(1+2z)+ 6q_0^2(4+77z+77z^2)\\
        &+ 7q_0(24+72z+(48 +5s_0)z^2)+3(24+4(12+s_0)z+ (24l_0+8s_0)z^2))\\
        &/(24 s_0+6l_0z-54q_0^5z+48s_0z+12j_0^3z^2-9q_0^6z^2+4s_0^2z^2-18q_0^4(2+11z)\\
        &+3q_0^2(-24+2(-12+5s_0)z+l_0z^2)+j_0^2(36+12(1+2q_0)z -23q_0^2z^2)\\
        &-3q_0^3(48+72z+s_0z^2)+6q_0(l_0z+s_0(4+15z))+j_0(72+66q_0^3z+12(6+s_0)z \\
        &-3l_0 z^2+24q_0^4z^2+6q_0^2(4+45z)+q_0(168+264z+11s_0z^2))))
            \label{eq:Hz_P22}
    \end{split}
\end{equation}

           \item $P_{32}$ equations of the luminosity distance and the Hubble rate:
\end{itemize}
\begin{equation}
\begin{split}
    d_L^{P_{32}(z)} & = \frac{cz}{H_0}\{-120-180s_0-156z-36l_0z-426s_0z-40z^2+80j_0^3z^2-30l_0z^2-135q_0^6z^2\\
    &-210s_0z^2+15s_0^2z^2-270q_0^5z(3+4z)+9q_0^4(-60+50z+63z^2)+2q_0^3(720+1767z\\
    &+887z^2)+3j_0^2(80+20(13+2q_0)z+(177+40q_0-60q_0^2)z^2)+6q_0^2[190+5(67\\
    &+9s_0)z+(125+3l_0+58s_0)z^2]6q_0[s_0(-30+4z+17z^2-2(20+(31+3l_0)z\\
    &+(9+4l_0)z^2)] +6j_0[-70+(-127+10s_0)z+45q_0^4z^2+(-47-2l_0+13s_0)z^2\\
    & +5q_0^3z(30+41z)-3q_0^2(-20+75z+69z^2)+2q_0(-115-274z+(-136+5s_0)z^2))]\}\\
    &\{3[-40-60s_0-32z-12l_0z-112s_0z 4z^2+40j_0^3z^2-4l_0z^2-135q_0^6z^2\\
    &-24s_0z^2+5s_0^2z^2-30q_0^5z(12+5z)3q_0^4(-60+160z+71z^2)\\
    &+j_0^2(80+20(11+4q_0)z +(57+20q_0-40q_0^2)z^2)+6q_0^3(80+188z+(44+\\
    &5s_0)z^2)+2q_0^2(190+20(13+3s_0)z+(46+6l_0+21s_0)z^2)+4q_0(20+\\
    &(16+3l_0)z+(2+l_0)z^2+s_0(15-17z-9z^2))2j_0(-70+2(-46+5s_0)z\\
    &+90q_0^4z^2+-16-2l_0+3s_0)z^2+15q_0^3z(12+5z+q_0^2(60-370z-141z^2)\\
    &+2q_0(-115-234z+2(-26+5s_0)z^2))]\}^{-1}
\end{split}
\end{equation}

\begin{equation}
   \begin{split}
       &H^{P_{32}}(z) = cH_0\{-540q_0^6-180q_0^3s_0+240s_0^2+36p_0q_0^2z-1620q_0^6z-1620q_0^7z-540q_0^3s_0z-180q_0^4s_0z\\
       &+720s_0^2z +660q_0s_0^2z+72p_0q_0^2z^2+72p_0q_0^3z^2-1620q_0^6z^2-3240q_0^7z^2-1215q_0^8z^2-12p_0s_0z^2\\
       &-540q_0^3s_0z^2-360q_0^4s_0z^2 +360q_0^5s_0z^2+720s_0^2z^2+1320q_0s_0^2z^2+315q_0^2s_0^2z^2+36p_0q_0^2z^3+72p_0q_0^3z^3\\
       &+18p_0q_0^4z^3 -540q_0^6z^3-1620q_0^7z^3-1215q_0^8z^3-135q_0^9z^3-12p_0s_0z^3-12p_0q_0s_0z^3-180q_0^3s_0z^3\\
       &-180q_0^4s_0z^3+360q_0^5s_0z^3+270q_0^6s_0z^3+240s_0^2z^3+660q_0s_0^2z^3+315q_0^2s_0^2z^3-45q_0^3s_0^2z^3-40s_0^3z^3\\
       &+15l_0^2z^2(1+z+q_0z)+60j_0^4z^2(10+(10+9q_0)z)+10j_0^3(72+12(18+19q_0)z+3(72+152q_0\\
       &+29q_0^2)z^2+(72+228q_0+87q_0^2-82q_0^3+3s_0)z^3)+6l_0(-12q_0^5z^3+12q_0^4z^2(1+z)+23q_0s_0z^2(1\\
       &+z)+66q_0^3z(1+z)^2+10s_0z(1+z)^2+3q_0^2(10+30z+30z^2+(10+s_0)z^3))+6j_0(45q_0^7z^3+\\
       &600q_0^6z^2(1+z)-105q_0^3s_0z^2(1+z)+780q_0^5z(1+z)^2-120q_0^4(-2-6z-6z^2+(-2+s_0)z^3)+\\
       &l_0(-30-2(45+28q_0)z+(-90-112q_0+41q_0^2)z^2+(-30-56q_0+41q_0^2+33q_0^3+10s_0)z^3)-\\
       &2z(1+z)(-25s_0^2z+3p_0(1+z))+2q_0^2z(-p_0z^2+95s_0(1+z)^2)+2q_0(10s_0^2z^3-7p_0z^2(1+z)+\\
       &55s_0(1+z)^3))+3j_0^2(75q_0^5z^3-1125q_0^4z^2(1+z)-1580q_0^3z(1 +z)^2+20q_0^2(-23-69z-69z^2+\\
       &(-23+10s_0)z^3)+2q_0z^2(-13l_0z+225s_0(1+z))+2z(50s_0(1+z)^2 -z(3p_0z+35l_0(1+z)))))\}\times\\
       &\times\{3(-180q_0^6-60q_0^3s_0+80s_0^2+12p_0q_0^2z-360q_0^6z-360q_0^7z-120q_0^3s_0z +160s_0^2z+140q_0s_0^2z+\\
       &80j_0^4z^2+5l_0^2z^2+12p_0q_0^2z^2+12p_0q_0^3z^2-180q_0^6z^2-360q_0^7z^2-135q_0^8z^2-4p_0s_0z^2-60q_0^3s_0z^2+\\
       &90q_0^5s_0z^2+80s_0^2z^2+140q_0s_0^2z^2+5q_0^2s_0^2z^2+40j_0^3(6+(12+13q_0)z+(6+13q_0+3q_0^2)z^2)+\\
       &l_0(-18q_0^4z^2+26q_0s_0z^2+72q_0^3z(1+z)+20s_0z(1+z)+60q_0^2(1+z)^2)-5j_0^2(8l_0z^2+95q_0^4z^2-\\
       &48q_0s_0z^2+224q_0^3z(1+z)-20s_0z(1+z)+92q_0^2(1+z)^2)+2j_0(225q_0^6z^2-115q_0^3s_0z^2+540q_0^5z(1\\
       &+z)+80q_0^2s_0z(1+z)+240q_0^4(1+z)^2-6z(p_0+p_0z-5s_0^2z)+l_0(-30-2(30+13q_0)z+(-30\\
       &-26q_0+37q_0^2)z^2) +2q_0(-4p_0z^2+55s_0(1+z)^2)))\}^{-1}
    \end{split}
        \label{eq:Hz_P32}
 \end{equation}



\end{document}

\endinput